# Scalability and Optimization Strategies for GPU Enhanced Neural Networks (GeNN)

Naresh Balaji[1], Esin Yavuz[2], Thomas Nowotny[2]

{T.Nowotny, E.Yavuz}@sussex.ac.uk, balajiravi25@gmail.com

[1]National Institute of Technology, Tiruchirappalli, India

[2]School of Engineering and Informatics, University of Sussex, Brighton, UK

*Abstract*: Simulation of spiking neural networks has been traditionally done on high-performance supercomputers or large-scale clusters. Utilizing the parallel nature of neural network computation algorithms, GeNN (GPU Enhanced Neural Network) provides a simulation environment that performs on General Purpose NVIDIA GPUs with a code generation based approach. GeNN allows the users to design and simulate neural networks by specifying the populations of neurons at different stages, their synapse connection densities and the model of individual neurons. In this report we describe work on how to scale synaptic weights based on the configuration of the user-defined network to ensure sufficient spiking and subsequent effective learning. We also discuss optimization strategies particular to GPU computing: sparse representation of synapse connections and occupancy based block-size determination.

## 1. Introduction

The computational performance of traditional single-core CPUs has steadily increased since its arrival, primarily due to the combination of increased clock frequency, process technology advancements and compiler optimizations. The clock frequency was pushed higher and higher, from 5 MHz to 3 GHz in the years from 1983 to 2002, until it reached a stall a few years ago [1] when the power consumption and dissipation of the transistors reached peaks that could not compensated by its cost [2]. The rising power consumption of CPUs with increase in the clock frequency paved way for an alternate strategy of increasing the computational cores rather than the clock cycles, thereby amortizing the heat dissipation to multiple points and still increasing the performance consistently. Intel termed this speed and power tradeoff as 'fundamental theorem of multi-core processors' [3]. Since then multi-core processers took center stage and have continued to evolve with performance metrics in terms of throughput rather than latency.

Artificial neural networks attempt to mimic the structure of the brain and replicate the processing and learning capabilities that the brain performs effortlessly. Although a large abstraction of the convoluted structure of the brain, ANNs are built only to incorporate the required bottom-up pathways for learning classification. Spiking neural network simulations are generally performed on dedicated hardware architectures or supercomputers. Usually, massive clusters of expensive computing hardware are used to avoid computational capabilities to pose as a bottleneck. As the biological neurons are performing in a massively parallel architecture, a replication of neural networks could benefit from comparable levels of parallelization. GeNN [4] is a neural network simulation environment using GPGPUs, hugely popularized by NVIDIA's parallel computing architecture termed CUDA, taking advantage of the inherent parallelism in neural network algorithms. Previous studies have shown a remarkable speed-up of GPUs and GeNN has recorded a speed boost of around 100x [5,6].

This article discusses the strategies involved in extending GeNN to different neuronal populations by suitably altering the synaptic conductance. Also discussed are the possible optimizations that could be incorporated by altering the representation of neurons and synapse elements in the constrained GPU memory-space and effectively exhausting the GPU latency by using optimum block and grid sizes.

## 2. Synaptic Conductance Scaling

Initially, modeling and simulation of neural networks using GeNN has been verified for consistency by implementing existing neural network models from the literature. In order to work with differing synapse connection density or population density of neuron groups, the existing synaptic conductance must be scaled to maintain a constant range of integrated synaptic current intakes at the post-synaptic neurons. But for determining the exact relationship and fitting in an algebraic function, various other factors need to be considered. Firstly, the scaling factors should maintain the arithmetic operands within the numeric range of the data type (single precision floating point is presently used in GeNN). When Hodgkin-Huxley model is used, the learning and evolution of the network are simulated by repeated calculations at discretized time steps. Although the time steps for computations can be minimized to values small enough to perform correct simulations, there exists a trade-off between the magnitude of time-steps and the real-time elapsed periods for the computation. When the time-steps are assigned to be large enough in allowing practically reasonable real-world simulation time, there are possibilities of the simulation overriding the self-stabilizing neurological models. Thus, an overflow in the precision arithmetic operands due to division by zeros or shooting up of numbers in magnitude results in 'not a number' (NANs). This contagiously affects other neurons due to their chained interconnectivities and results in erroneous values. Being highly dependent on the configuration of the preceding states, it is unlikely to arrive at a monotonic function that quantitative estimates the presence of NANs with the synaptic conductance. Therefore, it is necessary to perform a series of simulations in the likely range of the scaling factors for different pre-synaptic population. Secondly and probably the most important consideration for determining the scaling factor would be the quantitative level of spiking in the post-synaptic population. It is wiser to maintain the literature recommended percentage of spiking at each level of the network and extend it to different population densities. It is also essential to maintain the spiking densities within the prescribed range to ensure efficient learning at the subsequent stages.

We consider the effect of sparse vs. dense memory access patterns on the scaling and GPU vs. CPU execution speed-up. The representation of large matrices of neural synapses for computations could often lead to unnecessary usage of resources and processing time in the CPU. The storage is an even greater concern while working with GPUs because of the limited global memory and limited bandwidth of memory transfer between global and local memory. This is avoidable in many cases because of the sparse nature of connectivity and hence brings the need for a wiser utilization of resources. Substantial memory requirement reductions can be achieved by representing only the non-zero entries of the large sparse matrix, which gives rise to the next section.

## 3. GPU specific optimization strategies

The Compressed Row Storage format used in GeNN utilizes a structure of three arrays along with an integer storing the value of number of connections in the population sharing a synaptic group. The first array stores all the non-zero entries of the synapse connectivity as traversed along the pre-synaptic neuron indices. The second array stores the post-synaptic neuron indices as traversed along each of the pre-synaptic neuron indices. The last array stores the index of the latter array that starts each pre-synaptic neuron to its corresponding connections. The first and second arrays are of the same size as the number of non-zero entries of the otherwise dense matrix and the last array has size of pre-synaptic neuron population size. The equations (1) and (2) shows the memory requirements of sparse and dense representations, where nNZ, nPreSynN and nPostSynN denote the number of non-zero entries, number of pre-synaptic and post-synaptic neurons respectively. It can be seen that for considerably large populations of neurons, dense representation has a space complexity of $O(n^2)$ compared to $O(n+k)$ of sparse representation and the sparse representation almost always excels for realistic sparse models.

$$\text{Memory required for sparse representation} = 2 * nNZ + nPostSynN \qquad (1)$$
$$\text{Memory required for dense representation} = nPreSynN * nPostSynN \qquad (2)$$

The architecture of GPUs requires programmers to organize parallel threads into blocks and grids and handle their corresponding memory and processing resources. Although all the threads have access to global memory, threads within each block share the registers and shared memory of the SM they are dispatched to. Since the instructions issued by the scheduler are quantized as warps (a warp has 32 threads), the block size is almost always in multiples of 32 to avoid dead threads. The first choice of block size would be the maximum permitted by the hardware as it would mean larger independent block units and hence lesser inter-block memory shuffles. But since the SM resources are limited and are to be shared within the blocks in execution, the block size is to be decided by the resource usage of the block. The fundamental measure of performance in parallel architecture is the number of active warps per SM determined by 'occupancy', defined by NVIDIA as the ratio of resident warps to the maximum number of resident warps per SM. The occupancy is decided by one or more of the following four bottlenecks (i) maximum number of threads (ii) maximum number of blocks per SM (iii) shared memory and (iv) registers [7,8]. The block size is chosen to have sufficient occupancy to hide the memory transfer latency.

## 4. Materials and Methods

The CPU version was run on an iMac desktop with an Intel Core i5-4670 Quad-Core operating at 3.4GHz and having 8GB of memory. The GPU used was GeForce GT 755M with compute capability 3.0 and 1GB of global memory for the scalability analysis of network models. Simulations are run using GeNN and the offline analysis was conducted in Matlab (The Mathworks Inc., Natick, MA).

## 5. Results and Discussion

### 5.1 Scalability Analysis

The scaling of neuron populations was performed and tested for two different models of SNNs. The first one is an Izhikevich neuron [9] based network originally consisting of 1000 spiking cortical neurons with 1000 post-synaptic connections per neuron and the ratio of excitatory to inhibitory neuron population is 4 to 1. The post-synaptic connections per neurons are then varied from 100 to 1000 in steps of 50. The procedure discussed is followed for CPU and GPU version with both sparse and dense representations for verification purposes.

```
//Pseudo-code to optimize conductance scale factor
function [ coeff ] = gscale_optimise ( ref_nConn , ref_gScale )
        input_matrix := load file 'simulation_result.out'
        reference := find input_matrix at { 'nConn' == ref_nConn and 'gScale' == ref_gScale }
        //literature definition
        for ind = nConn.min : nConn.max,
                fringe := find input_matrix at { 'nConn' == ind },  //simulation results for each nConn
                output_matrix := append fringe at fringe['sumNANs']!=0 and min (abs (fringe['avgSpike'] –
                reference['avgSpike']  ))
                //finding gScale having closest average spike count to literature
        modelFuntion := fit output_matrix @(p,x) : p(1) / ( p(2)+x ) + p(3),  return p
```

Fig 1: Pseudo code showing the technique used to optimize the conductance scaling taking into consideration (a) the average spiking rate and (b) overflow of floating point arithmetic

We found that an inverse proportional relationship of the form $\text{gScale} = \frac{k1}{k2 + \text{nConn}} + k3$ between the scaling factor of synaptic conductance (gScale) and the population density of the pre-synaptic levels (nConn) was sufficient to provide the same spiking behaviour for increasing network sizes. The equation discussed is used to fit in the parameters for the gScale-nConn function, derived from the results of the simulation. Figure 1 shows the pseudo code that was used to find the optimal solution with the above considerations for synapse conductance scale. The values obtained by linear regression of the gscale function to the optimal values of scaling factor for a given number of connections is shown in Table 1. The fitting curves as well as the observed scaling values are shown in Figure 2. The difference between the scaling factors observed for sparse vs dense connectivity schemes and CPU vs GPU simulations was negligible. (Mean Absolute Percentage Error = 3.95%)

Table 1: Values of variables obtained by linear regression of the Izhikevich model

| k1 | k2 | k3 |
|---|---|---|
| $1.318 * 10^3$ | $1.099 * 10^2$ | $-2.800 * 10^{-1}$ |

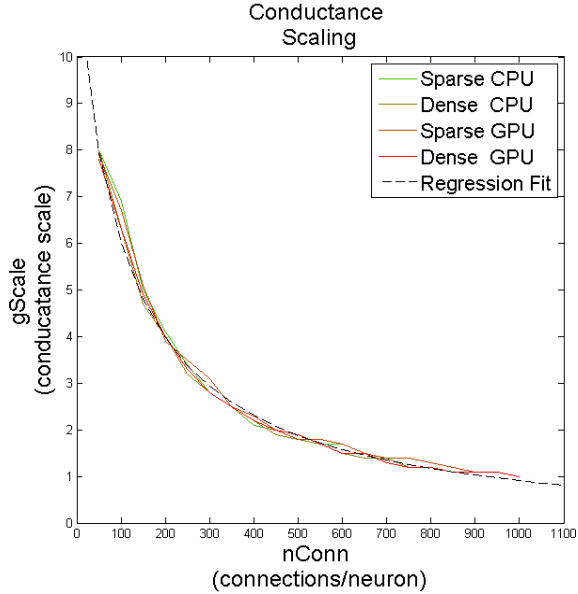

Fig 2: Simulation results in the Izhikevich neuron model between pre-synaptic neuron population connected to each post-synaptic neuron (nConn) and their corresponding synaptic-conductance scaling factors (gScale) performed in CPU and GPU with sparse and dense representations.

The other network considered is the model of the insect olfactory system [10] consisting of the input projection neurons (PN), lateral horn interneurons (LHI), intrinsic Kenyon cells (KC) and extrinsic Kenyon cells detection neurons (DN) of the mushroom body (MB) lobes. The population of PNs corresponding to the PN-KC and PN-LHI synapses is varied to determine the underlying synaptic conductance function. The experiment is performed with 1000 KCs and 100 DNs with 20 and 40 LHIs for later verification. (Mean Absolute Percentage Error for PN-LHI = 71.4% and for PN-KC = 16.1%)

Table 2: Values of variables obtained by linear regression of the insect olfaction model

| Synapse | k1 | k2 | k3 |
|---|---|---|---|
| PN-KC | $1.118 * 10^{-1}$ | 9.810 | $4.972 * 10^{-5}$ |
| PN-LHI | $1.354 * 10^3$ | $-6.338$ | $1.672 * 10^{-3}$ |

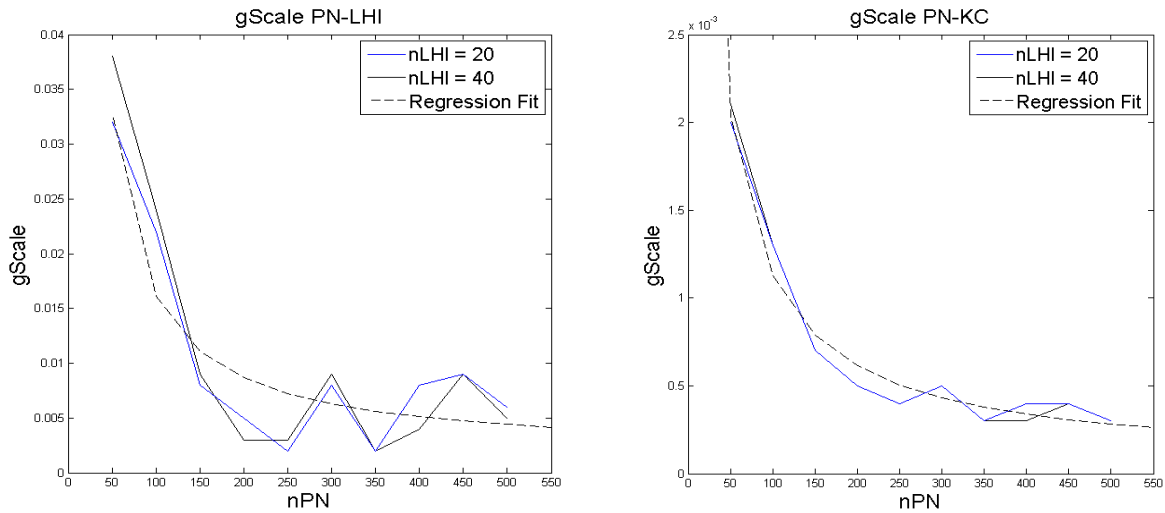

Fig 3: Simulation results in the olfactory mushroom body neuron model between pre-synaptic neuron population connected to each post-synaptic neuron (nConn) and their corresponding synaptic-conductance scaling factors (gScale) performed in with 20 and 40 LHI neurons.

Non-linear Regression over the scaling factors for a generic function and also their consistency over different configurations are shown in Table 2 and optimal scaling factors for a given number of PNs and the regression curve for PN-LHI and PN-KC synapse populations are shown in Figure 3. The resulting equation represents a function similar to $x*y = c$ with a shifted origin. The equation can now be rewritten as $(gScale - k3) * (nConn + k2) = k1$

## 4. Discussion

We performed scalability tests in order to compare the scaling of neuronal network models to ensure their correct function. First we performed a parameter exploration of the conductance scaling factor in order to maintain the same spiking behavior for different network sizes. Linear regression of the scaling factor for the Izhikevich model was very successful, while insect olfaction model required a nonlinear fitting and resulted in a less accurate fitting. This is probably due to the increased complexity of the scaling problem as a result of a more complex network with more nonlinear interactions in the insect olfaction model compared to the Izhikevich model. Another explanation is the non-deterministic nature of the Poisson neurons used in the insect olfaction model, which may result in a higher variability in the gscale values observed in Figure 3.

## 5. Acknowledgments

GeNN is open-source, free and available to download at genn.sourceforge.net. This work was performed as part of an internship of Naresh Balaji at the University of Sussex. Naresh Balaji would like to thank his family for their support and encouragement.